%% file: main.tex
\documentclass[prb,superscriptaddress,showpacs,twocolumn,amsmath,amssymb]{revtex4-1}
\input{Header_main}

\newcommand{\out}[1]{}

\newcommand{\eref}[1]{Eq.~(\ref{#1})}
\newcommand{\fref}[1]{Fig.~\ref{#1}}

\newcommand{\sref}[1]{Section~\ref{#1}}
\newcommand{\aref}[1]{\ref{#1}}

\newcommand{\vek}[1]{ \hbox{\textbf #1}}
\newcommand{\svek}{\mathbf}
\newcommand{\pr}{\ensuremath{^\prime}}

\newcommand{\cc}{%
        \ensuremath{c^\dag}   } 
\newcommand{\ca}{%
        \ensuremath{c^{\phantom{\dag}}}} 

\renewcommand{\Im}{\text{Im}}
\renewcommand{\Re}{\text{Re}}

\begin{document}

\title{Anisotropy of electronic correlations:\\
On the applicability of local theories to layered materials
}

\author{B.\ Klebel}
\affiliation{Institute of Solid State Physics, TU Wien, A-1040 Vienna, Austria}
\author{T.\ Sch\"afer}
\affiliation{Coll{\`e}ge de France, 11 place Marcelin Berthelot, 75005 Paris, France}
\affiliation{CPHT, CNRS, {\'E}cole Polytechnique, IP Paris, F-91128 Palaiseau, France}
\author{A.\ Toschi}
\affiliation{Institute of Solid State Physics, TU Wien, A-1040 Vienna, Austria}
\author{J.\ M.\ Tomczak}
\email{tomczak.jm@gmail.com}
\affiliation{Institute of Solid State Physics, TU Wien, A-1040 Vienna, Austria}

\begin{abstract}
Besides the chemical constituents, it is the lattice geometry that controls the most important material properties.
In many interesting compounds, the arrangement of elements leads to pronounced anisotropies, which reflect into a varying degree of quasi two-dimensionality of their low-energy excitations.
Here, we start by classifying important families of correlated materials according to a simple measure for the tetragonal anisotropy of their {\it ab initio} electronic (band) structure.
Second, we investigate the impact of a progressively large anisotropy in driving the non-locality of many-body effects.
To this end, we tune the Hubbard model from isotropic cubic in three dimensions to the two-dimensional limit and analyze it using the dynamical vertex approximation. For sufficiently isotropic hoppings, we find the self-energy to be well separable into a static non-local and a dynamical local contribution. 
While the latter could potentially be obtained from dynamical mean-field approaches, we find the former to be non-negligible in all cases.
Further, by increasing the model-anisotropy, we quantify the degree of quasi two-dimensionality which causes this ``space-time separation'' to break down. 
Our systematic analysis  improves the general understanding of electronic correlations in anisotropic materials, heterostructures and ultra-thin films, and provides useful guidance for future realistic studies.\\
\end{abstract}

\maketitle

\section{Introduction} 

\begin{figure}
            {\includegraphics[width = 1.05\linewidth] {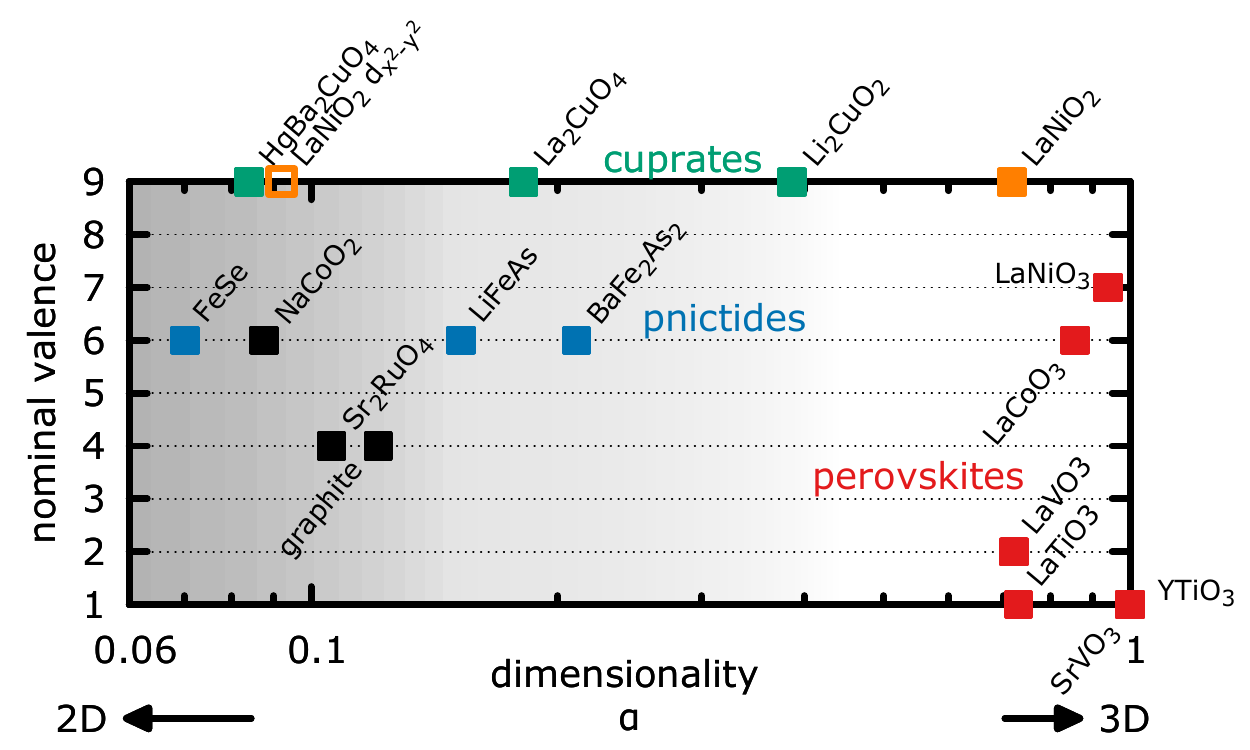}} 
            \caption{Overview of correlated materials classified by their dimensionality $\alpha$ (see \protect\eref{alpha}, \protect\sref{abinitio}) and the nominal valence of the transition-metal ions:  $\alpha=1$ for the cubic case in 3D, $\alpha=0$ corresponds to a planar 2D system. 
            Full (open) symbols indicate $\alpha$ when the low-energy electronic structure uses the full $d$-shell (the $d_{x^2-y^2}$ orbital) of the transition metal, see also Fig.~\protect\ref{fig:survey2}.
            We also include the $sp^3$-system graphite\protect\cite{Chung2002}.}
            \label{fig:survey}
        \end{figure}

Some of the most intensely studied condensed matter systems are 
highly anisotropic, see \fref{fig:survey}: cuprates, pnictides, ruthenates, cobaltates, graphene, transition-metal dichalcogenides, and ultra-thin oxide films.

Indeed, with spherical symmetry surrendered to translational invariance, the crystal-structure of periodic solids may trigger substantially anisotropic phenomena:
The coordination geometry of chemical constituents
can induce crystal-fields, the lifting of orbital degeneracies, and 
the directional dependence of transfer integrals. 
In layered compounds or geometrically engineered ultra-thin films, 
the latter effects conspire to produce low-energy dispersions
that are---to a degree---confined to a two-dimensional plane.
In \fref{fig:survey} we quantify such anisotropy for a number of prominent correlated materials
according to a measure $\alpha$ of their electronic structure, which we introduce below.

The effectively reduced dimensionality in the above mentioned one-particle ingredients
has significant consequences for many-body effects\cite{Aoki1992,Blundell2001,Larkin2008}:
First, low dimensions are the dominion of non-local  fluctuations in space: due to the low-coordination of the
lattice geometries, 
the physics of the system is strongly dependent on the specific spatial configuration realized at each step of its time-evolution.
As a result, in comparison to the 3D isotropic case, ordering instabilities are typically suppressed in quasi-2D materials or fully obviated in purely 2D systems (cf.\ Mermin-Wagner theorem\cite{PhysRevLett.17.1133} or Kosterlitz-Thouless transitions\cite{Kosterlitz1973}) by strong spatial fluctuations. The latter are reflected in significant enhancements of the corresponding susceptibilities, which  affect large regions of the parameter space in 2D, while in 3D these are typically confined to the proximity of the actual phase transitions\cite{RevModPhys.90.025003}. 
Second, these increased  (two-particle) fluctuations may affect one-particle spectral properties, as dictated by the Schwinger-Dyson equation of motion:
Corresponding renormalization effects, such as static energy shifts, quasi-particle effective weights and lifetimes may acquire strong non-local variations. The most prominent example is certainly the pseudogap regime in doped cuprates\cite{Timusk_1999,Norman2005,RevModPhys.78.17,keimer_quantum_2015},
but momentum-selective coherence and quasi-particle weights can in fact originate from various non-local fluctuations of spin, charge, or orbital degrees of freedom\cite{Matthias_SVOultra}.
Accordingly, pseudogap physics has been evidenced in a number of correlated materials, e.g.,
iron pnictides\cite{Xu2011,PhysRevLett.109.027006,PhysRevLett.109.037002,PhysRevB.89.045101} and chalcogenides\cite{PhysRevLett.111.217002}, iridates\cite{Kim187}, and (layered) nickelates\cite{PhysRevLett.106.027001}.%
\footnote{%
Further, there is indirect (=non-spectral), evidence for pseudogap physics in quasi-2D organic charge-transfer salts
(see, e.g., Ref.~\onlinecite{PhysRevB.80.054505} and references therein).
}

The most successful {\it local} approach for correlated electrons is dynamical
mean-field theory (DMFT)\cite{bible}, which can be combined with density functional theory, DFT+DMFT\cite{RevModPhys.78.865}, for {\it ab initio} material calculations. 
Exact in $D=\infty$\cite{PhysRevLett.62.324}, DMFT is empirically found to be
reasonable in three dimensions--- except in the vicinity of a
(second order) phase transition (for the reasons stated above).
However, recently it was shown\cite{jmt_dga3d}
that even in the presence of strong non-local (anti-ferromagnetic) fluctuations,
the {\it dynamical} part of many-body renormalizations (to linear order the quasi-particle weight) remains essentially local in 3D, while 
a notable variation within the Brillouin zone is engendered for {\it static} components of the self-energy.
As a result, the self-energy verifies---to a good approximation---a space-time separation\cite{jmt_dga3d}:
\begin{equation}
    \Sigma(\vek{k},\omega)=\Sigma_{\text{static}}(\vek{k})+\Sigma_{\text{local}}(\omega).
    \label{eqn:separation}
\end{equation}

In this work we explore and quantify the limits of this approximation:
We monitor the momentum-dependence of many-body renormalizations 
of the doped Hubbard model as a continuous function 
of the tetragonal anisotropy in the one-particle hopping---from the cubic case in 3D to the square lattice in 2D.
We find that for an anisotropy 
smaller than {roughly} one half ($\alpha<1/2$), a local approximation to the dynamical self-energy becomes inadequate. 
In conjunction with the survey of
electronic-structure anisotropies (\fref{fig:survey}),
our results hence provide guidance for future first principle investigations of layered correlated systems:  We  establish a rule-of-thumb when 
the use of techniques beyond DFT+DMFT\cite{Tomczak2017review,RevModPhys.90.025003,Anna_ADGA,Lechermann2017} becomes a prerequisite
for {reliable} {\it ab initio} calculations. 

The paper is organized as follows: We detail the employed
methodology in \sref{method}.
In \sref{results} we present and discuss our results: \sref{abinitioresults}
is devoted to the {\it ab inito} classification of materials according to the anisotropy of their electronic structure. In \sref{modelsection}
we study the Hubbard model for the range of anisotropy covered by the materials in \sref{abinitioresults}.
We conclude in \sref{perspective} with a synthesis of the materials classification and our many-body findings.

\section{Methods}\label{method}
\subsection{Ab initio calculations}\label{abinitio}

For the {\it ab initio} survey of materials, \fref{fig:survey}, we (i) performed (non-spinpolarized) density functional calculations (using the PBE functional) with WIEN2k\cite{wien2k}, (ii)
constructed maximally localized Wannier functions\cite{RevModPhys.84.1419} for the $d$-orbitals of the transition metal relevant for the low-energy band-structure with wannier90\cite{wannier90} via wien2wannier\cite{wien2wannier}, and (iii) analyzed the hopping amplitudes of the one-particle Hamiltonian in real-space: $H=\sum_{\tau\tau\pr, LL\pr, RR\pr,\sigma}H^{\tau\tau\pr}_{LL\pr}(R-R\pr)\cc_{\tau\pr L\pr R\pr \sigma}\ca_{\tau LR\sigma}$. Here, $\tau$ indexes transition-metal atoms in the unit-cell, $L$
the $d$-orbitals (or a subset of them, see below), $R$ the unitcell, and $\sigma$ the spin.
The measure $\alpha$ is then defined as the ratio of the in-plane and out-of-plane hopping amplitudes between  nearest-neighbour transition-metal ions:
\begin{equation}
    \alpha = \frac{\max_{\tau_\perp,R_\perp}\sum_{LL\pr}|H^{\tau\tau_\perp}_{LL\pr}(R_\perp)|}{\max_{\tau_\parallel,R_\parallel}\sum_{LL\pr}|H^{\tau\tau_\parallel}_{LL\pr}(R_\parallel)|}
    \label{alpha}
\end{equation}
To the denominator (nominator) contribute all 
orbital combinations of hopping elements between transition-metal ions
that are nearest-neighbours in the $ab$-plane (in the $c$-direction). Taking the maximum, e.g., for the numerator, over $(\tau_\perp,R_\perp)\ne(\tau,R)$ assures selecting transition-metal ions that are nearest-neighbours.
For example,
$R_\perp=0$ ($R_\parallel=0$) for an intra-cell hopping between nearest neighbours $\tau\ne\tau_\perp$ ($\tau\ne\tau_\parallel$);
while $R_\perp=a_z\, \hat{\vek{e}}_z$ ($R_\parallel=a_x\,\hat{\vek{e}}_x$, or, if $a_y<a_x$, $R_\parallel=a_y\,\hat{\vek{e}}_y$) 
with $\tau=\tau_\perp$ ($\tau=\tau_\parallel$) indicates the hopping between the same atom but in adjacent unit-cells translated by the lattice constant $a_x$ ($a_y$) in the direction of the  respective unit-vector $\hat{\vek{e}}_x$ ($\hat{\vek{e}}_y$). 
The anisotropy as described by \eref{alpha} is clearly too simple to account for {\it global} material trends.
In particular, $\alpha$ largely depends on the orbital-subspace chosen to represent the pertinent low-energy dispersions (this will be detailed below for the high-$T_c$ parent compounds). However, we believe $\alpha$ to be a good indicator for trends {\it within} a given family of compounds. 
As such, the anisotropy proxy of \eref{alpha} could even serve
as descriptor in high-throughput studies that target the optimization of some property through, e.g., chemical pressure via iso-valent substitutions\cite{jmt_radialKI}.

\subsection{Many-body calculations}\label{model}

\subsubsection{The model}\label{modelintro}

\begin{figure}[!t]
            \centering
            {\includegraphics[clip=true,trim=0 225 450 0, width = .75\linewidth] {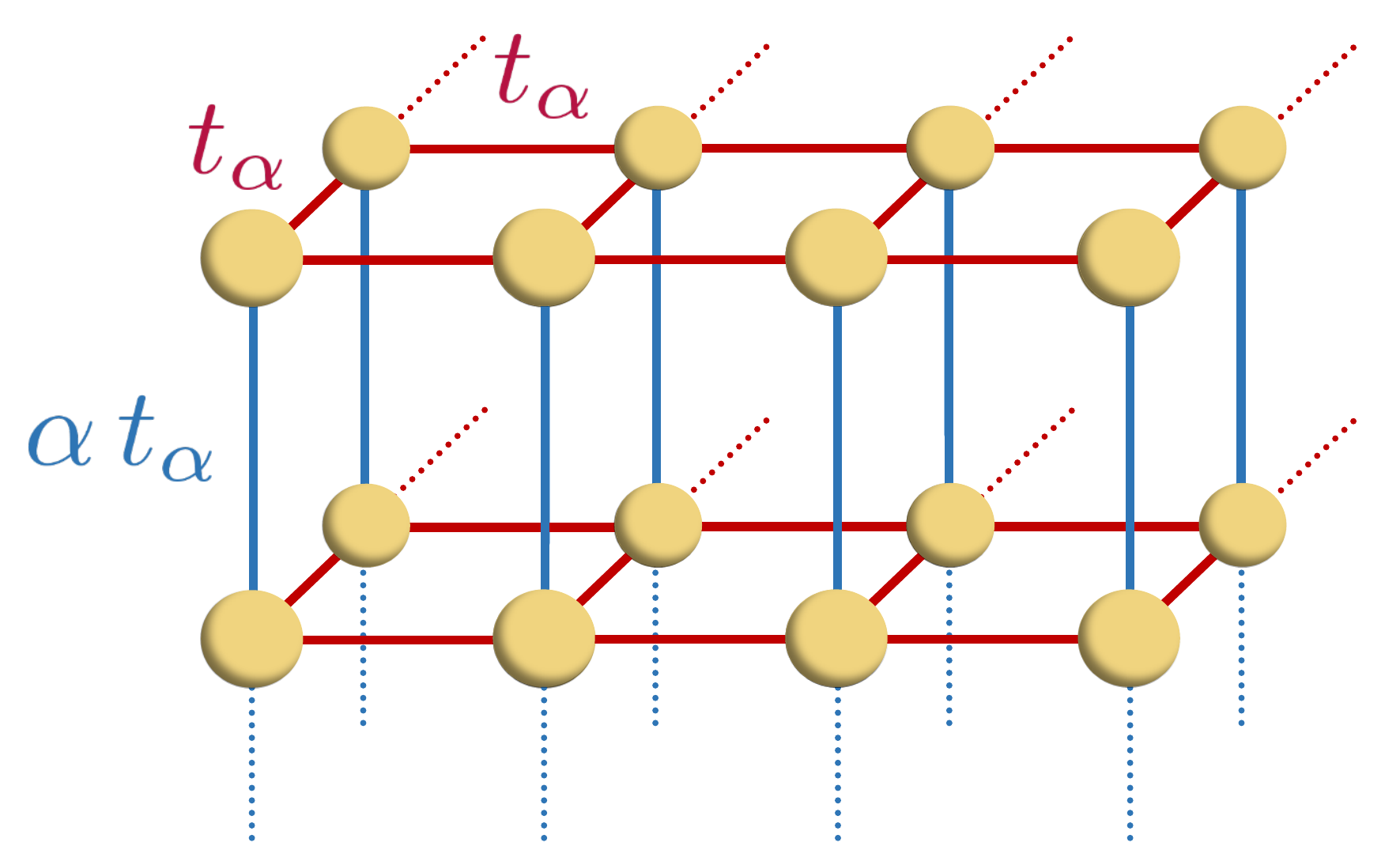}} 
            \caption{The tetragonal lattice: In the $xy$-plane, electrons hop with an amplitude $t_\alpha$, while the electron transfer in $z$ direction is smaller by a factor $0\leq\alpha\leq1$.}
            \label{fig:hub}
\end{figure}

We consider the one-band Hubbard model (in the usual notation)
\begin{eqnarray}
    H&=&\sum_{\svek{k}\sigma} \epsilon_\svek{k}^{\phantom{^\dagger}}
     \cc_{\svek{k}\sigma}\ca_{\svek{k}\sigma}+U\sum_i n_{i\uparrow} n_{i\downarrow}\label{ham}
\end{eqnarray}
on a tetragonal lattice (see \fref{fig:hub}):
\begin{eqnarray}
    \epsilon_{\svek{k}}&=&-2t_\alpha 
     \left[ \cos(k_xa_x)+\cos(k_ya_y)+\alpha\cdot\cos(k_za_z)\right]
     \label{dispersion}
\end{eqnarray}
where $k_i\in[0,2\pi/a_i)$ and we set $a_i=1$ ($i=x,y,z$).
The parameter $\alpha\in[0,1]$ is the one-band equivalent of \eref{alpha} as it scales the hopping amplitude in the $z$-direction:
It allows to continuously tune the system between the 3D cubic case ($\alpha=1$) and the
2D square lattice ($\alpha=0$).
Such change in the anisotropy  directly affects, per construction, the overall kinetic energy and, hence---for a fixed interaction---the degree of electronic correlations in the system. 
As this trivial effect would hide the most interesting trends emerging from our dimensional investigation, we chose to keep the kinetic energy essentially independent of $\alpha$. To this end, we adjust the overall hopping amplitude $t_\alpha$ by requiring
the second moment of the density of states to be constant and fix our energy units by imposing: 
\begin{equation}
    \int_\infty^{-\infty} \epsilon^2 N(\epsilon)\,d\epsilon \stackrel{!}{=}\frac{1}{4},
\end{equation}
where $N(\epsilon)=\sum_{\svek{k}}\delta(\epsilon-\epsilon_\svek{k})$. 
This requirement
yields the analytical expression
\begin{equation}
    t_{\alpha}=\frac{1}{2\sqrt{4+2\alpha^2}}.
\end{equation}
\fref{DOS} in App.~\ref{app1} displays the corresponding density of states (DOS) varying continuously from three to two dimensions (see also Ref.~\onlinecite{Vollhardt_Jerusalem}). 
    The DOS of a hypercubic lattice in infinite dimensions is---as a result of the central limit theorem---a Gaussian. In 3D, the DOS is still reasonably close to that Gaussian, yet with a flattened top.
    Going to 2D, however, two significant changes appear: 
    The flat top narrows into a single peak (a logarithmic divergence: the so-called van-Hove singularity), whereas the sides develop \enquote{knees} (which become van-Hove singularities in 1D). 
    We note that the 
    {movement} of the knees towards the band-edge and the narrowing of the central plateau is somewhat linear with the change in dimensionality, while the increase in height of the narrowing plateau starts slow, and only speeds up significantly below 
    $\alpha\leq 0.5$, i.e.\ for an effective dimension D $\le 2.5$.
    Note that, in this work, we neglect all complications introduced by next-nearest neighbour hoppings\cite{PhysRevLett.87.047003} or ionic potentials\cite{PhysRevB.98.235105}.
     
\subsubsection{Dynamical vertex approximation}
In order to be able to reliably estimate how non-local fluctuations on top of DMFT affect the space time-separation of the self-energy in Eq.~(\ref{eqn:separation}), we use a diagrammatic extension \cite{RevModPhys.90.025003} of DMFT, the dynamical vertex approximation (D$\Gamma$A\cite{toschi_dynamical_2007}) in its most used  ladder-implementation with Moriya corrections \cite{katanin_comparing_2009,rohringer_impact_2016}. In contrast to quantum cluster methods \cite{maier_quantum_2005}, it includes temporal and non-local correlations at all lengths scales on an equal footing. For this reason, D$\Gamma$A has been often exploited to describe classical \cite{rohringer_critical_2011,delRe2019} and quantum phase transitions \cite{PhysRevLett.119.046402, schafer_pam_2019}, as well as the associated fluctuations. For further technical details on the D$\Gamma$A calculation see App.~\ref{app:dga}. 

After the calculation of the momentum-dependent self-energy on the Matsubara axis $\Sigma(\mathbf{k},i\omega_{n})$, 
we adapt the chemical potential $\mu$ in order to {recover} the system's filling,
due to the lack of full self-consistency of ladder-D$\Gamma$A with Moriya corrections\cite{katanin_comparing_2009,rohringer_impact_2016}.

Throughout the paper, we focus on many-body renormalizations near the Fermi level. To this end, we  perform a Taylor series expansion, specifically 
\begin{eqnarray}
\Sigma(\vek{k},\omega)&=&\Re\Sigma(\vek{k},\omega=0)+(1-1/Z(\vek{k}))\omega\nonumber \\
&&+i\Gamma(\vek{k})(\omega^2+\pi^2T^2)+\cdots, 
\end{eqnarray}
where $\gamma(\vek{k})=-\Im\Sigma(\vek{k},\omega=0)=\Gamma(\vek{k})\pi^2T^2+\mathcal{O}(T^4)$ is the scattering rate and $Z(\vek{k})$ the quasi-particle weight---provided the system realizes a Fermi-liquid.
These quantities are extracted from the self-energy on the Matsubara axis, $\Sigma(\mathbf{k},i\omega_{n})$, as follows:
\begin{eqnarray}
\Re \Sigma(\mathbf{k},\omega=0)&=&\Re\Sigma(\mathbf{k},i\omega_{n})\big\vert_{\omega_{n}\rightarrow{0}}\label{eqS0}\\
 \gamma_{\mathbf{k}}&=&-\Im\Sigma(\mathbf{k},i\omega_{n})\big\vert_{\omega_{n}\rightarrow{0}}\label{eqg}\\
 Z_{\mathbf{k}}&=&\left[1-\frac{\partial\text{Im }\Sigma(\mathbf{k},i\omega_{n})}{\partial\omega_{n}}\bigg\vert_{\omega_{n}\rightarrow{0}}\right]^{-1}\label{eqZ}
\end{eqnarray}

To obtain stable results for our analyzed quantities (see next subsection), we performed and averaged over polynomial fits up to order {10} before the extrapolation $i\omega_n\!\rightarrow\!0$.

In all our calculations we use $U\!=\!2$ in units of $t_\alpha$,
placing us---at zero doping---in the intermediate-coupling regime, yielding the highest magnetic transition temperature in 3D \cite{rohringer_critical_2011}. In this work, we will restrict ourselves to non-local correlations stemming from magnetism in the paramagnetic regime (paramagnons).

\subsubsection{Analysis tools}

To analyse our results, we introduce the following measures
 of non-locality\cite{jmt_dga3d,jmt_pnict} for the  expansion coefficients defined above in Eqs.~(\ref{eqS0})-(\ref{eqZ}), $a(\vek{k})=\Re\Sigma(\vek{k},\omega=0)$, $Z(\vek{k})$, and $\gamma(\vek{k})$: 
\begin{enumerate}[(\roman{enumi})]
    \item 
the standard deviation 
                \begin{equation}  \Delta_{\kvec} a(\kvec) = \sqrt{1/N_{\kvec} \sum_{\kvec} {\left| a(\kvec) - a_{loc}  \right|}^2   }    \label{measure-1} \end{equation}
with respect to the local (Brillouin-zone average) value
%
    $a_{loc}=1/N_\svek{k}\sum_\svek{k}a(\vek{k})$.
%
\item the standard deviation on the Fermi surface (FS), i.e.\ $\vek{k}=\vek{k}_F$:
    \begin{equation} \Delta_{\kvec}^{FS} a(\kvec) = \sqrt{1/N_{\kvec_{F}} \sum_{\kvec_{F}} {\left| a(\kvec) - a_{FS}  \right|}^2   }, \label{measure-2}
    \end{equation}
with $a_{FS}=1/N_{\svek{k}_F}\sum_{\svek{k}_F}a(\vek{k})$.
Irrespective of potentially large scattering rates that may invalidate the quasi-particle picture, we determine the Fermi surface as momenta $\vek{k}_F$ that verify the quasi-particle equation, $\det (\mu-\epsilon_{\svek{k}_F}-\Re\Sigma(\vek{k}_F,0))=0$.
\item the maximum absolute difference on the Fermi surface          
    \begin{equation} \delta_{\kvec}^{FS} a(\kvec) = \left| {\text{max}}_{\svek{k}_F}(a(\kvec_F)) - \text{min}_{\svek{k}_F}(a(\kvec_F)) \right|.    \label{measure-3}  \end{equation}
                
                \end{enumerate}

\section{Results \& Discussion}\label{results}
\subsection{Dimensionality of materials: a classification}\label{abinitioresults}

In \fref{fig:survey} we classify some important correlated materials according to the dimensionality of their electronic (band) structure (following \eref{alpha}) and their nominal valences. As low-energy model we 
{choose} the full shell of $d$-orbitals of the respective transition metal.

        \begin{figure}
            \centering
            \centerline{\includegraphics[clip=true,trim=0 0 0 40,width = 1\linewidth] {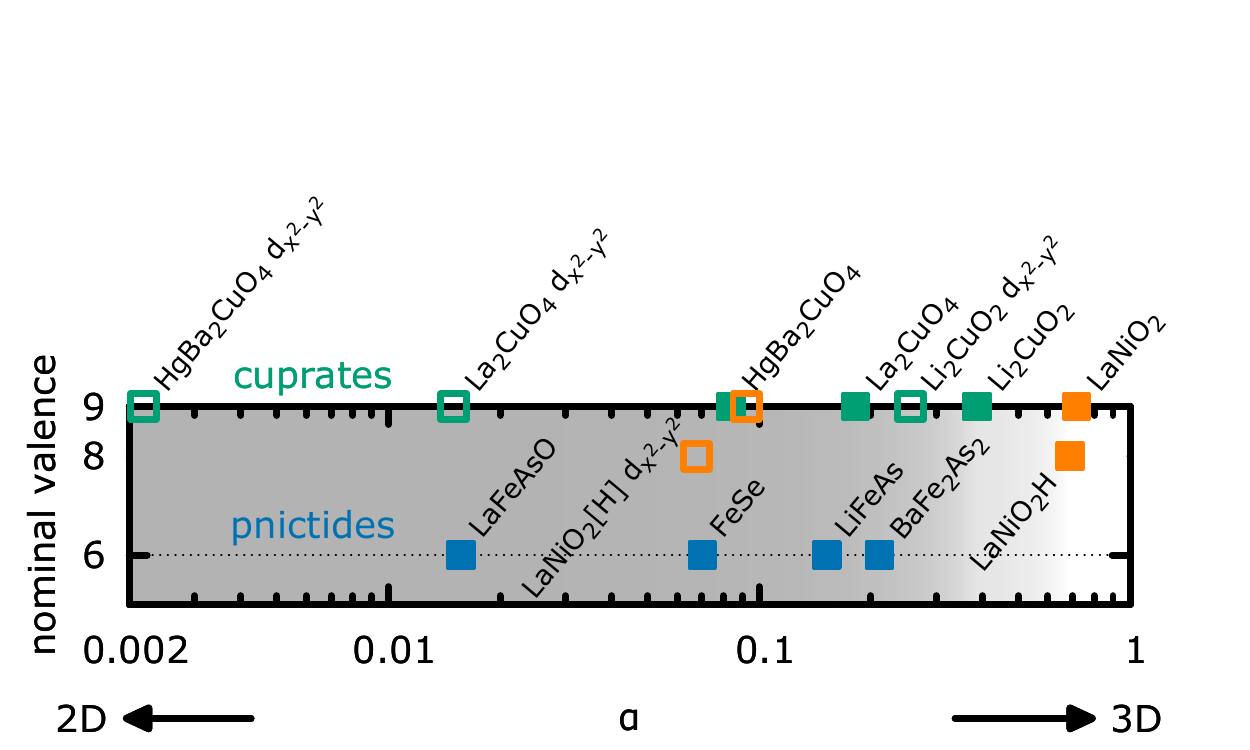}} 
            \caption{Dependence of the dimensionality $\alpha$ on the chosen low-energy orbital subspace: solid symbols correspond to the values from \fref{fig:survey} that use the full $d$-shell of the transition metal, open symbols use a single-orbital (Cu/Ni-$d_{x^2-y^2}$) modeling.}
            \label{fig:survey2}
        \end{figure}

The most 3D-like materials included are perovskite oxides, such as titanates, vanadates and nickelates of the form $AM$O$_3$\cite{imada} (group 2/3 element or rare-earth {\it A}, transition metal {\it M}).
In fact, the perfect cubic perovskite structure---realized, e.g., in SrVO$_3$---has $\alpha=1$
by construction.
A dramatic dimensional reduction can, however, be achieved by deploying these materials
in ultra-thin films. Such geometric constraints may lead to a
reduction of out-of-plane hopping \cite{PhysRevLett.104.147601} (i.e.\ $\alpha<1$) 
as well as orbital polarizations \cite{PhysRevLett.114.246401}, to the extent
that, e.g., SrVO$_3$ or CaVO$_3$-thin films undergo a metal-insulator transition below a critical thickness \cite{PhysRevLett.104.147601,Kobayashi2017,McNally2019}.
A path to lower symmetries in bulk perovskites are 
distortions and tiltings of the oxygen octaheadra surrounding the transition metal $M$.
The resulting orthorhombic structures typically exhibit an increased degree of electronic correlations.
Indeed, for $d^1$-perovskites, the {ensuing} lifting of degeneracies quenches orbital fluctuations and triggers a Mott transition\cite{pavarini:176403,1367-2630-7-1-188}.
However, this physics is not driven by inter-orbital hybridizations
but dominated by changes in local crystal-fields, which do not directly affect the value of $\alpha$. Indeed, for the Mott insulator 
YTiO$_3$ $\alpha\approx 1$  despite pronounced deviations from a perfect cubic structure.

Dimensional effects captured by Eq.~(\ref{alpha}) are instead crucial in oxides of copper:
We find $\alpha\!\approx\!0.4$ for (non-superconducting) Li$_2$CuO$_2$, while the two- and one-layer high-$T_c$ parents compounds La$_2$CuO$_4$ and HgBa$_2$CuO$_4$\cite{barisic_demonstrating_2008}
have consecutively smaller $\alpha$. The quasi two-dimensional physics in cuprates seems to be
indelibly connected with the occurrence of high-temperature superconductivity under doping (see, e.g., Ref.~\onlinecite{Leggett2006}). 
Indeed, structural (and also chemical) variations will lead to trends in various electronic degrees of freedom, and
empirical connections between the latter and $T_c$ have been established. Among these (interlinked) indicators are
in-plane next-nearest neighbour hoppings \cite{PhysRevLett.87.047003}, the charge-transfer energy \cite{Weber_2012,PhysRevX.8.021038},
the $e_g$-splitting \cite{PhysRevLett.105.057003,PhysRevB.85.064501},
and the magnitude of in-plane magnetic exchange couplings 
\cite{Chang_LSCO}. 
Here, we note that, for the shown cuprates, the smaller $\alpha$, the larger $T_c$ at optimal doping.

Also in the case of the iron pnictides and chalcogenides (in their tetragonal structures), the dimensionality parameter $\alpha$ follows intuition:
The 122-family is the most isotropic, the 111 less so, and the 11-chalcogenide is the most 2D-like, see \fref{fig:survey}.
This trend neatly follows the magnitude of electronic correlations as monitored by, e.g., by the effective mass
\cite{Yin_pnictide} or the fluctuating
magnetic moment\cite{Yin_pnictide,PhysRevB.86.064411,watzenboeck2020}.%
\footnote{A notable outlier is LaFeAsO (see \fref{fig:survey2}), which in our measure, \eref{alpha}, is the most 2D-like pnictide considered. 
This finding is congruent with the fact that the Fe-Fe distance in this 1111-compound is larger by about 50\% than in the other pnictides considered. However, effective mass renormalizations in LaFeAsO\cite{PhysRevB.81.100502,doi:10.1143/JPSJ.77.093714} is comparable to BaFe$_2$As$_2$, the most 3D-like pnictide here. This deviation in the trend is ascribed to the lower value of the screened interaction in the $3d$-orbital manifold of the 1111-material\cite{Miyake2009}.}

Comparing the $\alpha$-values of the families of cuprates and pnictides in \fref{fig:survey}, one notices that---contrary to common belief--- the latter are overall as quasi two-dimensional as the former. As alluded to before, we stress that the measure $\alpha$ crucially depends on the choice of the orbital subset sought to represent the low-energy electronic structure. Comparisons of materials 
are therefore most adequate when they allow for a common orbital framework, thus, in particular, when the materials belong to the same family of compounds. Indeed, in \fref{fig:survey}, $\alpha$ is computed from the hoppings of the full $d$-shell of the transition metal. 
Contrary to the multi-orbital pnictides with nominal $d^6$-valence\cite{Yin_pnictide},
the low-energy physics of $d^9$ cuprates is mostly dominated by a single orbital: the $d_{x^2-y^2}$ (see, however, e.g., Refs.~\onlinecite{PhysRevLett.58.2794,PhysRevLett.105.057003,tabis_w_notitle_nodate}).
The smaller orbital space limits the possibility for other dimensional effects beyond the purely geometrical ones.
In \fref{fig:survey2} we therefore report $\alpha$-values for cuprates and the nickelate LaNiO$_2$\cite{LaNiO2}%
$^,$\footnote{
related to the recently discovered infinite-layer $d^9$ nickelate superconductor Nd$_{0.8}$Sr$_{0.2}$NiO$_2$\cite{Li2019,Wilson2019}. We also include LaNiO$_2$H, a $d^8$ Mott insulating\cite{Liang_LaNiO2} by-product in the synthesizing of LaNiO$_2$.
}
within said single-orbital framework:
While trends within the cuprates remain qualitatively unchanged, the whole family substantially moves towards the 2D-limit.
The anisotropy of LaNiO$_2$ surpasses 90\%, but is still small as compared to the high-$T_c$ parent compounds (making it tempting to suggest finding layered $d^9$ nickelates with smaller $\alpha$ to potentially increase $T_c$).

Having established the effective dimensionality of pertinent classes of correlated materials, we now investigate how the structure-driven electronic anisotropy 
affects the non-locality of many-electron renormalizations.
As anticipated, we do so in the context of a simple model, allowing us to identify global trends without complicating the picture with material specific degrees of freedom.

\subsection{The Hubbard model on a tetragonal lattice}\label{modelsection}

    \begin{figure}
            \centering
            \centerline{\includegraphics[width = 1\linewidth] {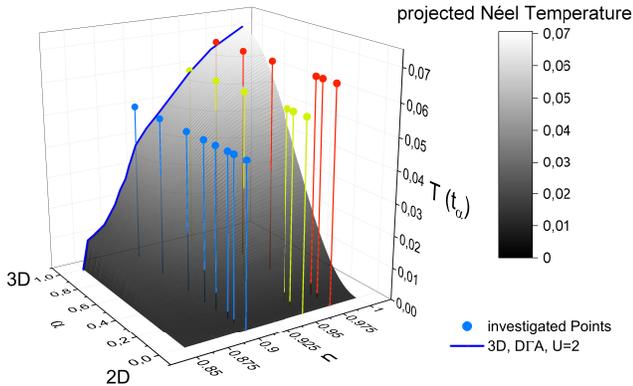}}  
            \caption{Phase diagram of the tetragonal Hubbard model (\protect\eref{ham}): temperature $T$ vs.\ doping $n$ and dimensionality $\alpha$. The blue curve corresponds to the N{\'e}el temperature $T_{N}$ in 3D ($\alpha=1$)\protect\cite{jmt_dga3d,PhysRevLett.119.046402}; the gray gradient is a guide to the eye
            that qualitatively estimates $T_{N}$ for $\alpha<1$ with a cosine function.
            Coloured pins indicate the points in parameter space chosen for our investigation: $(n,\beta)=(0.975,15)$ (red), $(0.95,17)$ (yellow), and $(0.9,20)$ (blue).
            }
            \label{parameter_space}
        \end{figure}

\subsubsection{Results \& Discussion}\label{modelresults}

\begin{figure*}
        \begin{subfigure}{0.8\textwidth}
            \centering
                \begin{subfigure}{0.27\textwidth}
                    \centerline{\includegraphics[clip=true,trim=0 0 0 0,width = 1\linewidth]{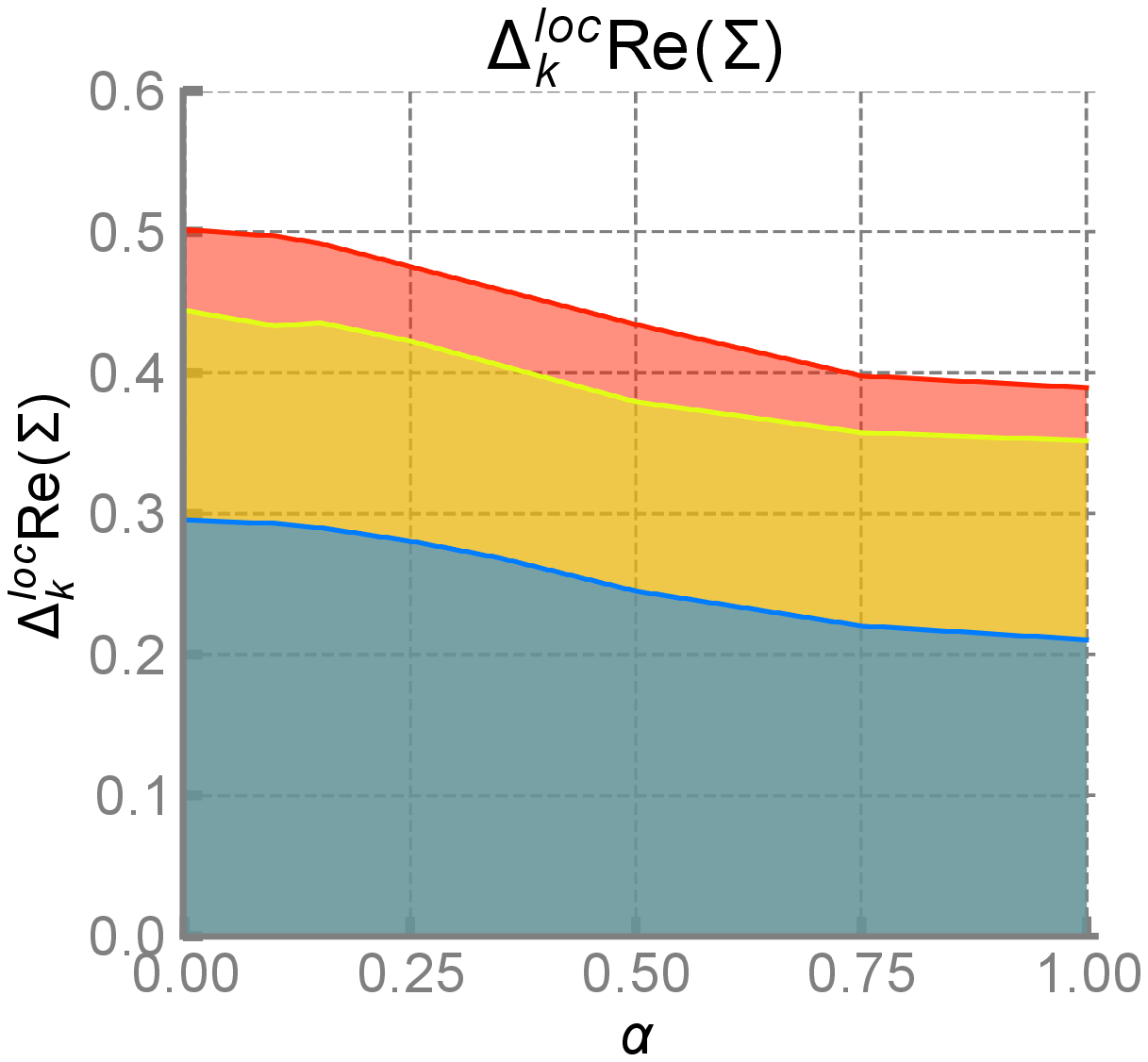}}
                    \caption{}
                    \label{fig:plt:res:resigma}
                \end{subfigure}
                \hfill
                \begin{subfigure}{0.27\textwidth}
                    \centerline{\includegraphics[clip=true,trim=0 100 0 100,width = 0.6\linewidth] {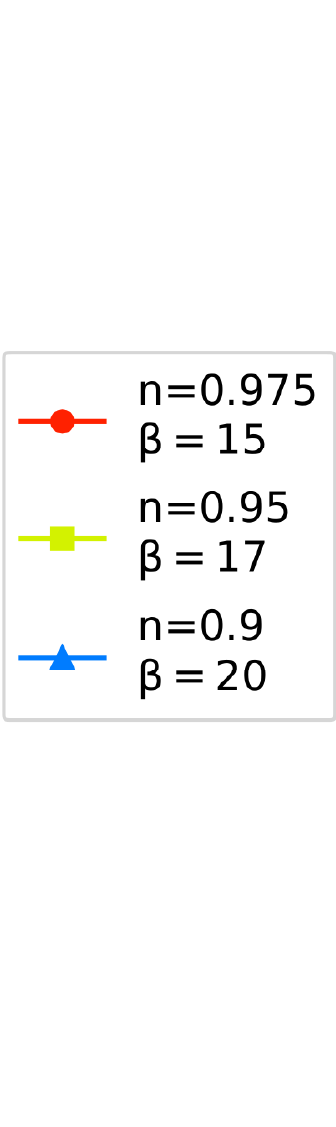}}  
                    \caption{}
                    \label{fig:plt:res:legend}  
                \end{subfigure}
                \hfill
                \begin{subfigure}{0.27\textwidth}
                    \centerline{\includegraphics[clip=true,trim=0 0 0 0,width = 1\linewidth]{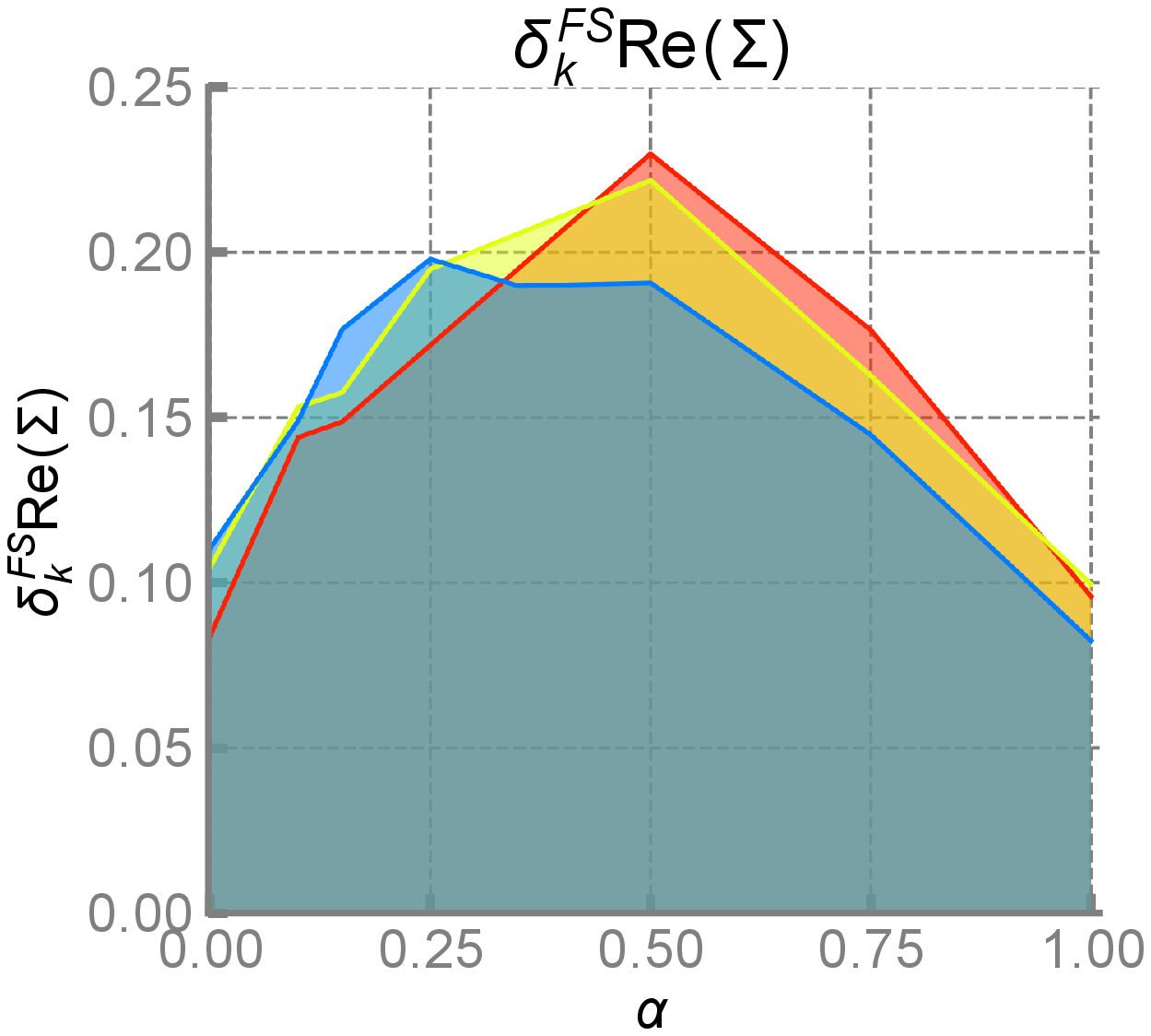}}
                    \caption{}
                    \label{fig:plt:res:resigma_minmax}  
                \end{subfigure}
                \vfill
                \end{subfigure}     
    
            \begin{subfigure}{0.8\textwidth}
            \centering
                \begin{subfigure}{0.27\textwidth}
                    \centerline{\includegraphics[width = 1\linewidth] {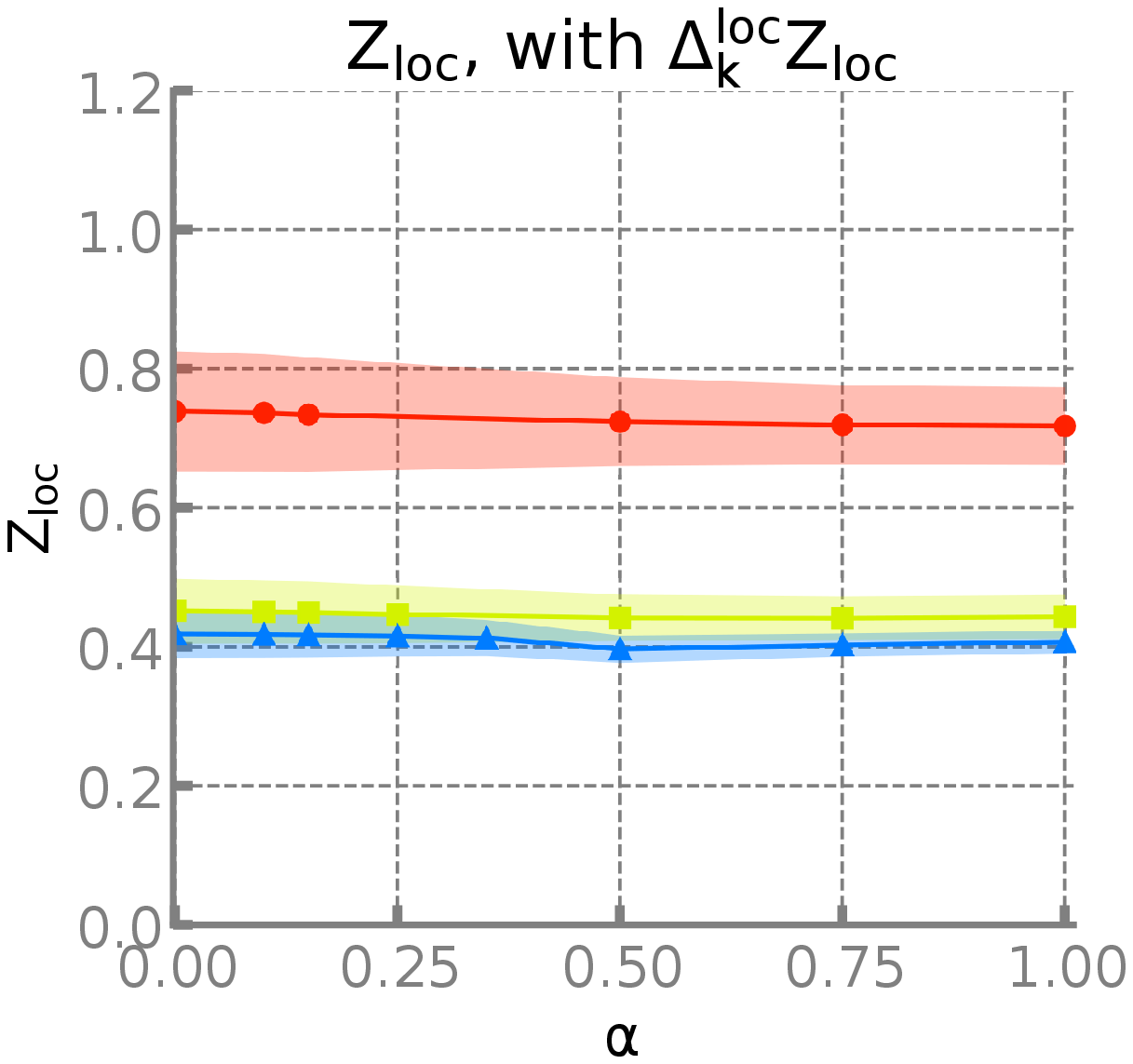}}
                    \caption{}
                    \label{fig:plt:res:Zalpha_BZ_varianz}
                \end{subfigure}
                \hfill
                \begin{subfigure}{0.27\textwidth}
                    \centerline{\includegraphics[width = 1\linewidth] {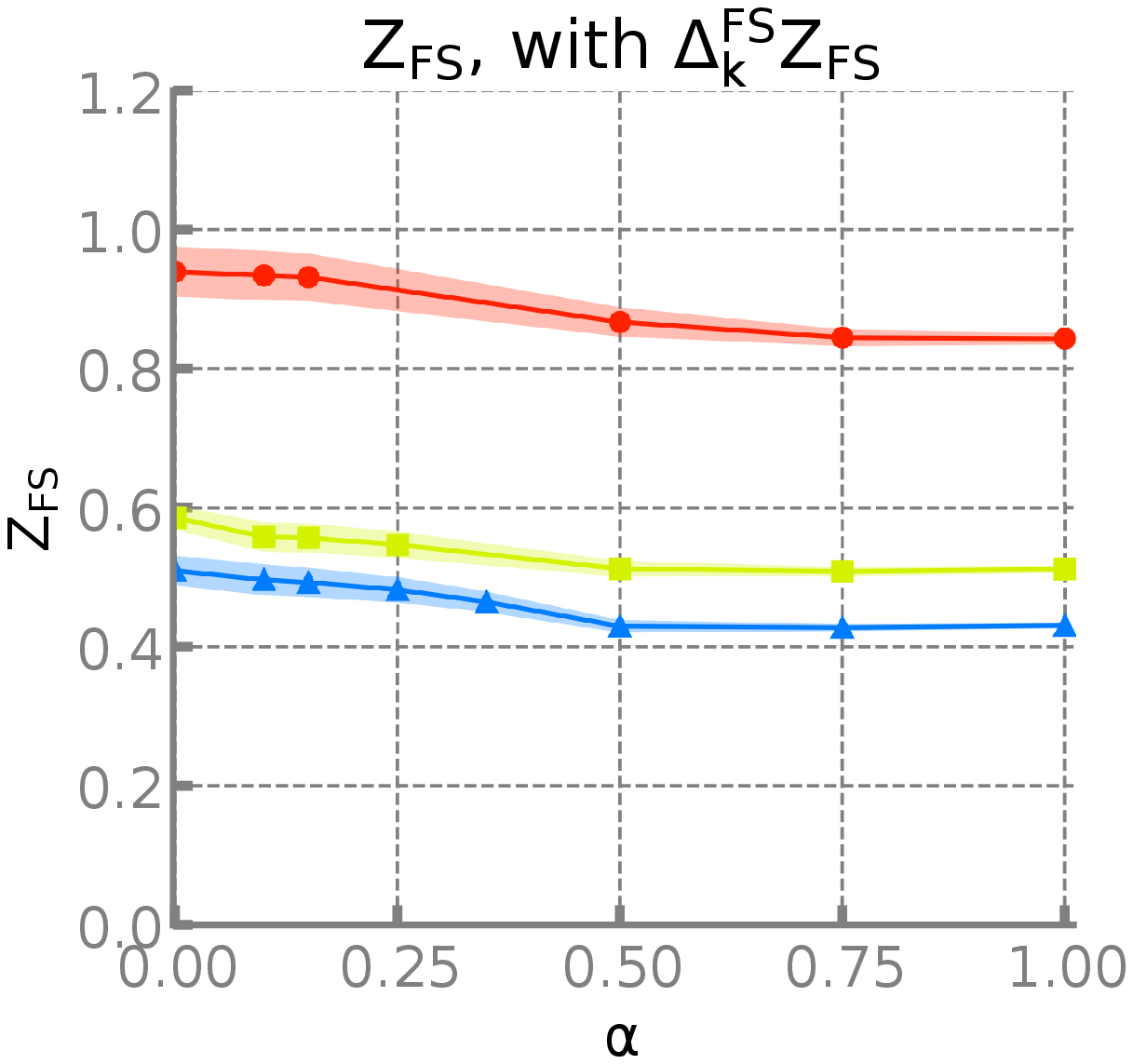}}
                    \caption{}
                    \label{fig:plt:res:Zalpha_FS_varianz}
                \end{subfigure}
                \hfill              
                \begin{subfigure}{0.27\textwidth}
                    \centerline{\includegraphics[width = 1\linewidth] {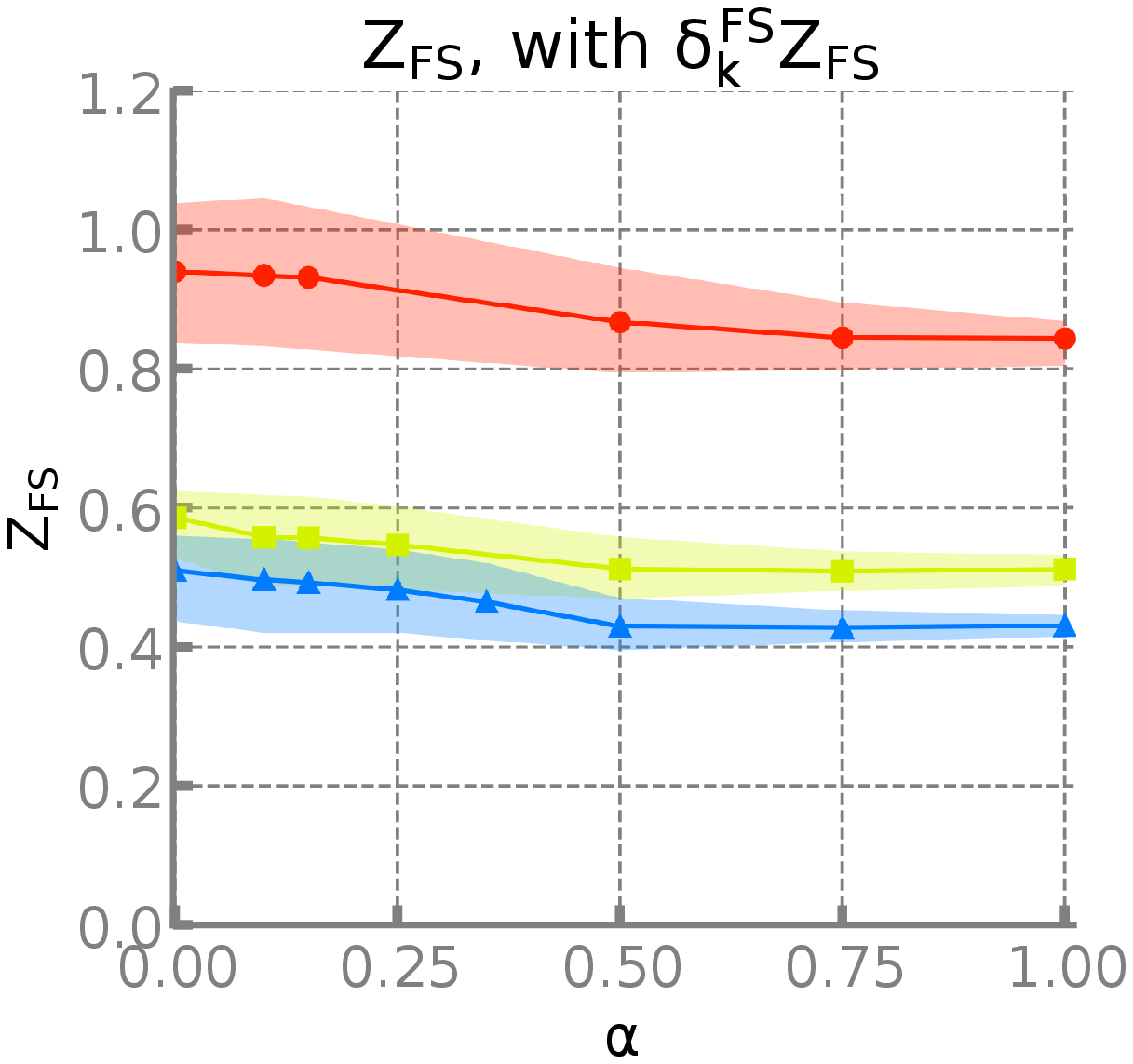}}
                    \caption{}
                    \label{fig:plt:res:Zalpha_FS_minmax}    
                \end{subfigure}         
                \vfill  
                
                \begin{subfigure}{0.27\textwidth}
                    \centerline{\includegraphics[width = 1\linewidth] {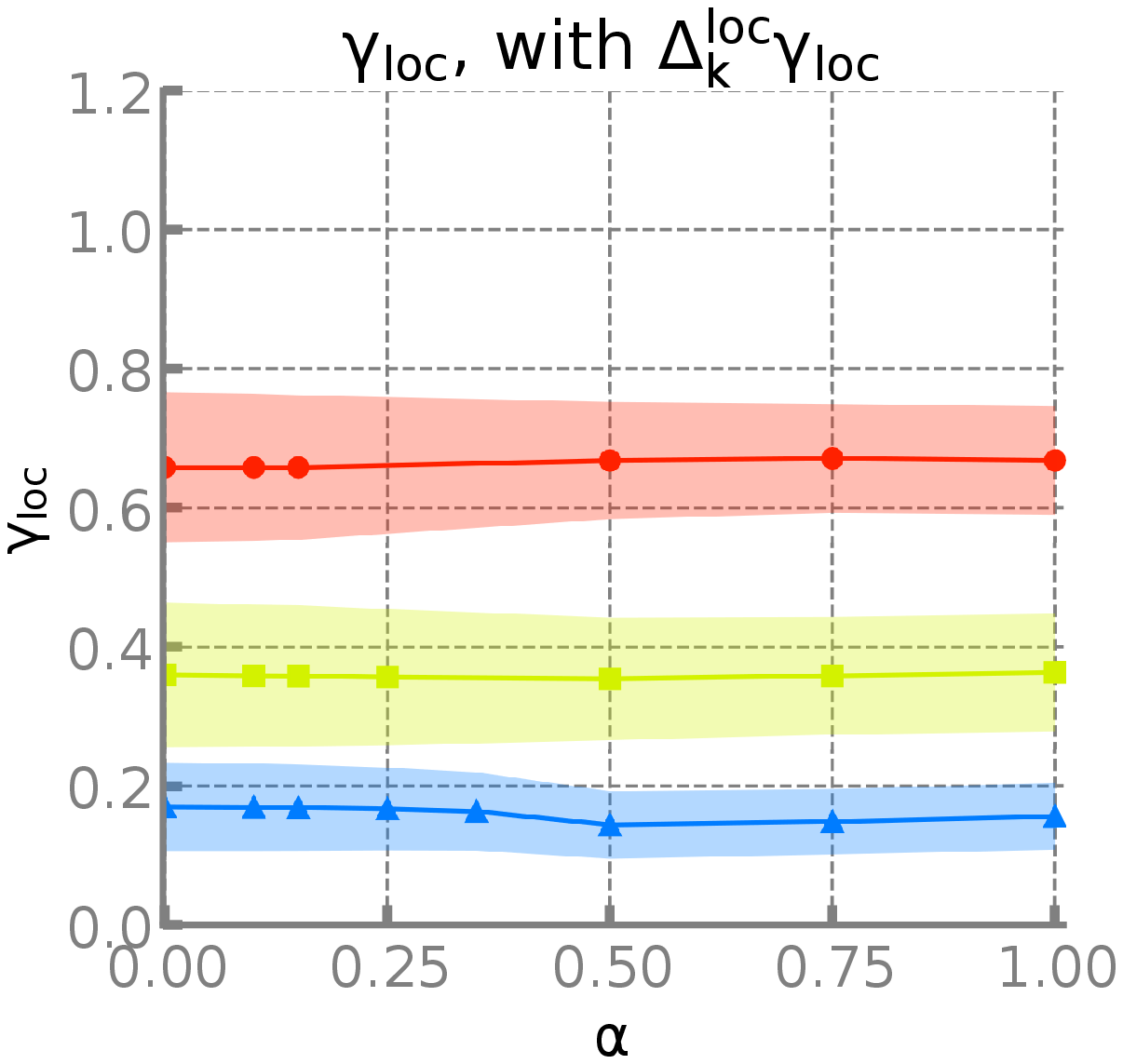}}
                    \caption{}
                    \label{fig:plt:res:gammaalpha_BZ_varianz}
                \end{subfigure}         
                \hfill
                \begin{subfigure}{0.27\textwidth}
                    \centerline{\includegraphics[width = 1\linewidth] {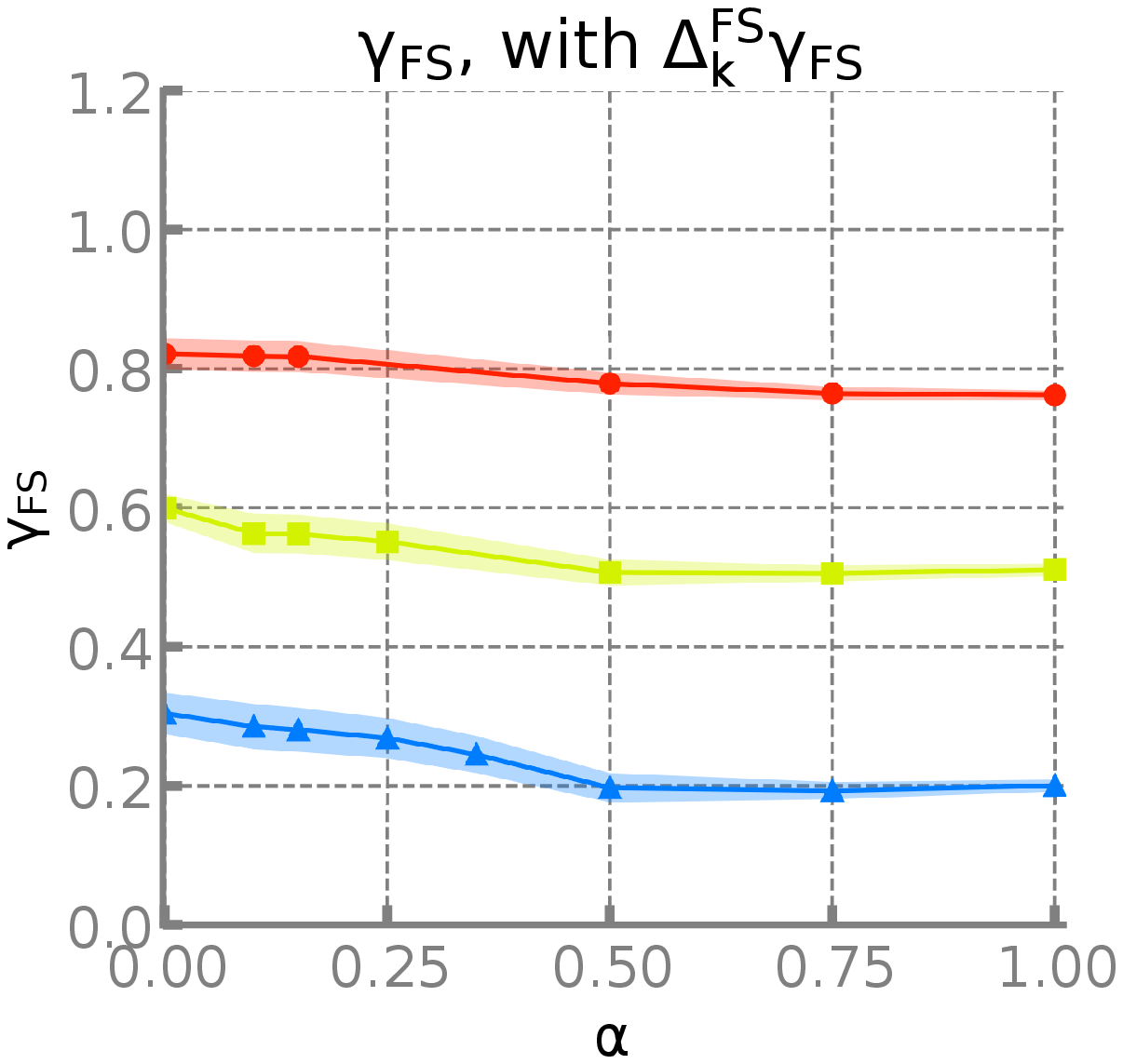}}
                    \caption{}
                    \label{fig:plt:res:gammaalpha_FS_varianz}
                \end{subfigure}
                \hfill
                \begin{subfigure}{0.27\textwidth}
                    \centerline{\includegraphics[width = 1\linewidth] {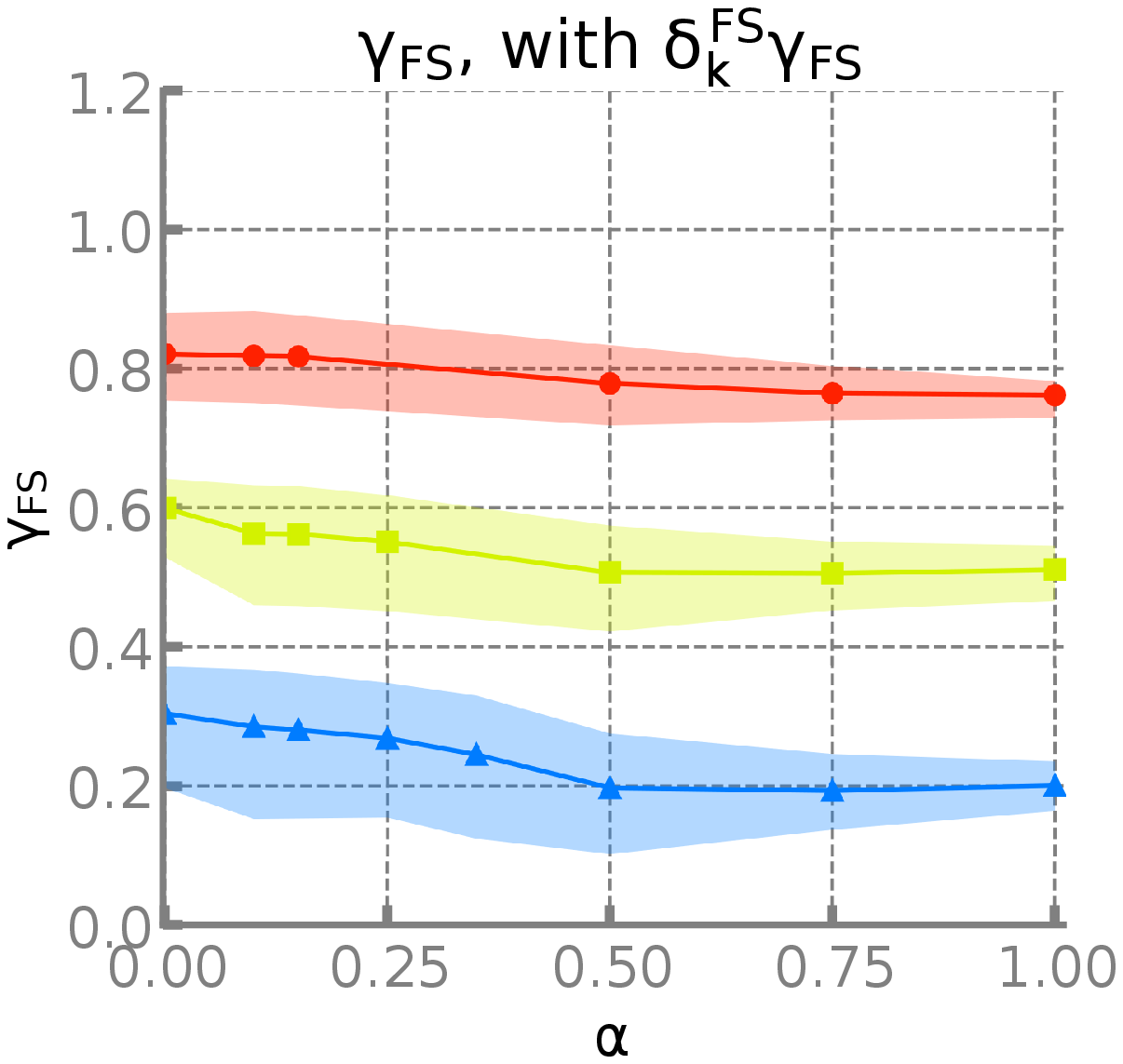}}
                    \caption{}
                    \label{fig:plt:res:gammaalpha_FS_minmax}    
                \end{subfigure} 
            \end{subfigure}
            \caption{Quasi-particle parameters and their momentum-dependence:
            $\Re\Sigma(\vek{k},\omega=0)$ (top), $Z(\vek{k})$ (middle), $\gamma(\vek{k})$ (bottom) analysed in terms of the measure
            $\Delta^{loc}_{\svek{k}}$ (Brillouin zone average, \protect\eref{measure-1}, left), 
            $\Delta^{FS}_{\svek{k}}$ (Fermi surface average, \protect\eref{measure-2}, middle)
            $\delta^{FS}_{\svek{k}}$ (maximal absolute difference in Fermi surface,\protect \eref{measure-3}, right) as a function of the dimensionality $\alpha$ for the parameters shown in panel (b).
             \DfsRe{} (not shown) is very similar to \minmaxRe{}.}
            \label{results_measures}
        \end{figure*}

We solve the Hamiltonian \eref{ham} for anisotropy parameters $\alpha$, dopings $n$, and temperatures $T=(k_B\beta)^{-1}$ as indicated in \fref{parameter_space}.%
\footnote{As a caveat to the trends described below, let us note that with decreasing $\alpha$, the system is located further and further away from the spin-ordered phase. Therefore, proximity effects are expected to be stronger close to 3D than for (quasi-)2D, which should be considered when discussing the overall effect of changes in the dimensionality.
}
Essentially, in this work we introduce with the dimensionality $\alpha$ a new axis to the temperature vs.\ doping phase diagram of Ref.~\onlinecite{jmt_dga3d}.\footnote{
For $\alpha<1$ magnetic transition temperatures $T_{N}$ displayed as shadings are only guides to the eye. The $T_{N}$-curve for $\alpha\!=\!1$ has been reproduced from Refs.~\onlinecite{jmt_dga3d,PhysRevLett.119.046402}.}
\paragraph{Measures of non-locality.}
\fref{results_measures} displays the non-locality of $\Re\Sigma(\vek{k},\omega=0)$ [top], $Z(\vek{k})$ [middle], $\gamma(\vek{k})$ [bottom] according to the measures  Eq.~(\ref{measure-1}) [left], Eq.~(\ref{measure-2}) [middle], and Eq.~(\ref{measure-3}) [right] for the chosen regimes. For both  $Z(\vek{k})$ and $\gamma(\vek{k})$ the momentum-variation is shown as a shaded region around the local value.

We first discuss the differences in these 
{\sl local} quantities: 
The  scattering rate $\gamma_{loc}$ is larger the smaller the doping, i.e.\ it is largest in direct proximity to half-filling ($n=1.0$), where an essentially insulating behavior has been evidenced by previous D$\Gamma$A studies\cite{rohringer_critical_2011,rohringer_impact_2016,RevModPhys.90.025003}.
At any doping, the scattering rate is consistently larger when averaged over the Fermi surface rather than the full Brillouin zone (compare panels (g) and (h)), as it is expected in the presence of strong non-local spin fluctuations. 
At least at small doping the large scattering rate $\gamma$ impedes the interpretation of $Z$ as the quasi-particle weight. For instance, for $n=0.975$ and $\beta=15$ (data marked in red in \fref{results_measures}), we find  $\gamma_{loc}\approx 65\%$ 
of the 2D half-bandwidth, clearly invalidating the Fermi liquid picture.%
\footnote{Indeed, we previously noted in 3D\cite{jmt_dga3d} that in this regime $\gamma$ involves corrections\cite{PhysRevB.68.155113} to Fermi liquid theory in its temperature dependence, 
as neither DMFT nor D$\Gamma$A verify a $T^2$ behaviour.}
As a consequence, the fact that we find for this set of parameters the largest $Z_{loc}\approx 0.7$ does not contradict the fact of being the most correlated regime considered.

Next, we turn to the $\alpha$-dependence of the {\sl local} expansion coefficients:
When averaged over the entire Brillouin zone, both the $Z$-factor and the scattering rate $\gamma$ hardly change when the dimensionality is reduced.
However, when limiting the momentum-average to the Fermi surface, i.e.\ to the region where low-energy excitations are actually present, there is a significant upturn in both $Z$ and $\gamma$ (panels (e) and (h)) when $\alpha$ drops below $0.5$.
This trend is consistent with the behaviour of the underlying DOS (shown in the Appendix's \fref{DOS}, and discussed in \sref{modelintro}; see also the supplementary material\cite{supps}) at low energies:
Indeed, the density at the Fermi level drastically increases when the effective dimension drops below $2.5$ (i.e.\ $\alpha\leq 0.5$), providing carriers available for electronic scattering and thus driving up the collision rate.
In particular at our largest dopings, this change in the trend is almost cusp-like.

Finally, we address the momentum-dependence of the quasi-particle parameters.
As a function of $\alpha$ the standard deviation throughout the Brillouin zone (following \eref{measure-1}) of both $Z$ and $\gamma$ slightly increases from 3D to 2D. Overall, however, their momentum-dependence is moderate in this measure. The momentum-variation of static shifts $\Re\Sigma(\vek{k},\omega=0)$, on the other hand,
is already large in 3D,
in accordance with previous results\cite{jmt_dga3d}.
When reducing $\alpha$, this spread increases continuously for all investigated dopings and temperatures. In the most correlated case ($n=0.975$), the standard deviation $\Delta^{loc}_{\svek{k}}\Re\Sigma(\vek{k},\omega=0)$
grows from 
31\% (3D) to above 50\% (2D) of the respective half-bandwidth (cf.\ \fref{DOS} and \onlinecite{supps}).
Static non-local renormalizations are indeed non-negligible for all studied regimes.

The variation on the Fermi surface, measured via \eref{measure-2} 
reveals, instead, a notable enhancement towards lower dimensionality:
While the standard deviation of $Z$ and $\gamma$ (panels (e) and (h)) is virtually zero in 3D, it acquires consistently growing finite values for smaller $\alpha$. The spread in both quantities is, however, significantly smaller than when averaged over the whole Brillouin zone (panels (d) and (g)).

Since the averaging of the measures \eref{measure-1} and \eref{measure-2} might obfuscate a strong momentum variation carried by only a few $\vek{k}$-points, we plot in panels (c), (f), (i) the maximal spread on the Fermi surface according to \eref{measure-3}. The increased momentum resolution of this analysis reveals a much larger variation than in the $\vek{k}$-averaged data for $Z$ and $\gamma$, indicating that, as a matter of fact, the momentum dependence must be driven by {\sl small areas} of the Fermi surface, where large deviations from the local average are found.

As to the trend with doping, we must note that the behavior of the momentum-spread of the scattering rate on the FS is inverted with respect to that on the whole Brillouin zone: the largest momentum-spread is realized for the largest doping. More quantitatively,  we find that the difference between the maximal and the minimal scattering rate surpasses 100\% of the local value $\gamma^{FS}$ on the Fermi surface below $\alpha=0.5$.

        \begin{figure*}
            \begin{subfigure}{0.28\textwidth}
                \centerline{\includegraphics[clip=true,trim=0 40 0 50,width = 0.9\linewidth]{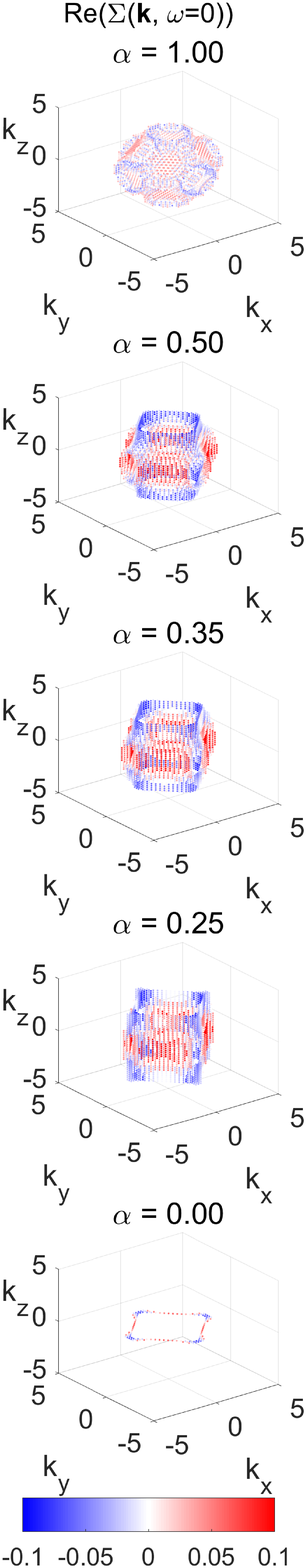}} 
                \caption{}
                \label{fig:plt:res:FS_Real}
            \end{subfigure}         
            \hfill
            \begin{subfigure}{0.28\textwidth}
                \centerline{\includegraphics[clip=true,trim=0 40 0 50,width = 0.9\linewidth]{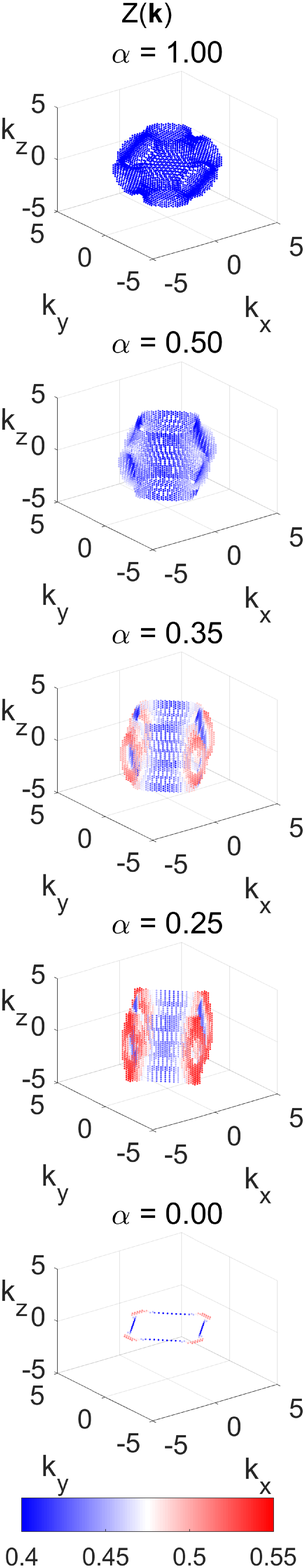}} 
                \caption{}
                \label{fig:plt:res:FS_Z}
            \end{subfigure}
            \hfill
            \begin{subfigure}{0.28\textwidth}
                \centerline{\includegraphics[clip=true,trim=0 40 0 50,width = 0.9\linewidth]{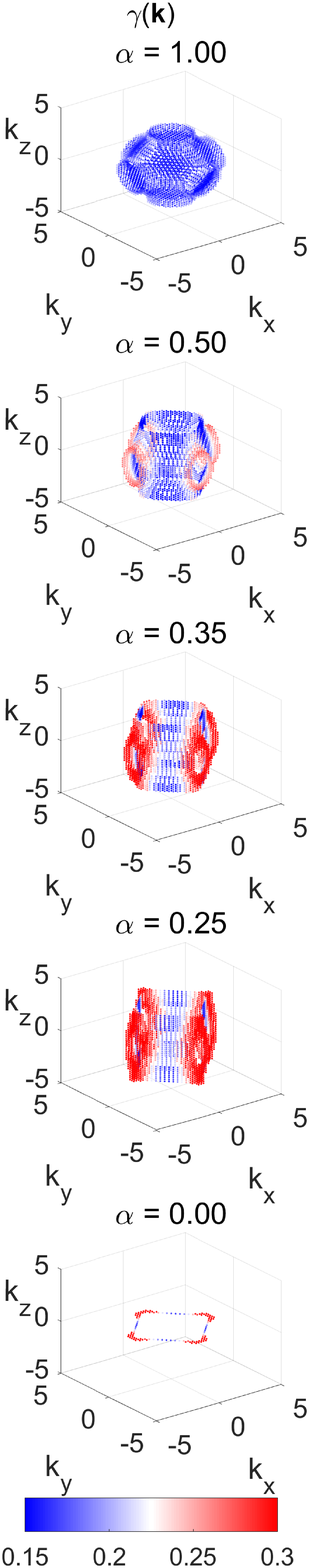}} %
                \caption{}
                \label{fig:plt:res:FS_Gamma}
            \end{subfigure}
            \caption{Renormalizations on the Fermi surface: $\Re\left[\Sigma(\kvec{}_F,\omega=0)-1/N_{\svek{k}_F}\sum_{\svek{k}_F}\Sigma(\vek{k}_F,\omega=0)\right]$, $Z(\vek{k}_F)$,  and $\gamma{}(\vek{k}_F)$ for $n=0.90$, $\beta=20$ and representative values of $\alpha$ on the respective D$\Gamma$A Fermi surface. 
            \label{3DFS} }
        \end{figure*}

Also for $\Re\Sigma(\vek{k},\omega=0)$, the $\alpha$-dependence is quite different when looking at peak values (panel (c)): instead of a continuous increase for shrinking $\alpha$, the variation peaks at intermediate dimensionality and diminishes towards 2D.
We speculate that the following contributes to the reversal of the trend:
$\Re\Sigma(\vek{k},\omega=0)$ describes the deformation of the Fermi surface with respect to the local DMFT starting point. For a metallic solution (and low enough $T$), this deformation is subject to
conserving the Fermi surface volume (Luttinger's theorem) and C$_4$ symmetry in the ab-plane. Now, the decreasing dispersion along the c-axis introduces a strong geometric constraint that further impedes deformations (see also Appendix \aref{sapp1}).
Furthermore  we find, empirically, that 
\begin{equation}
    \left.\hat{\vek{e}}_{\svek{k}}\cdot\nabla_{\svek{k}}\Re\Sigma(\vek{k},\omega=0) / \hat{\vek{e}}_{\svek{k}}\cdot \nabla_{\svek{k}}\epsilon_{\svek{k}}\right|_{\svek{k}=\svek{k}_F}>0,\label{ineq}
    \end{equation}
    suggesting that the sign of the Fermi velocity perpendicular to the Fermi surface determines the direction of deformation, further limiting its freedom.

\paragraph{Renormalizations on the Fermi surface.}
The two measures of non-locality on the Fermi surface---\eref{measure-2}: FS-average, \eref{measure-3}: maximal absolute difference---yield notably different results, suggesting largely inhomogenous renormalizations.
Here we investigate this further by explicitly plotting in \fref{3DFS}
$\Re\Sigma(\vek{k},\omega=0)$ (left column, with respect to the Fermi surface average, $1/N_{\svek{k}_F}\sum_{\svek{k}_F}\Re\Sigma(\vek{k}_F,\omega=0)$), $Z(\vek{k})$ (middle), and $\gamma(\vek{k})$ (right) on the Fermi surface for the set of parameters given by  $n=0.9$, $\beta=20$ and various $\alpha$ (rows).

In 3D ($\alpha=1$, top row in \fref{3DFS}), the values of the $Z$-factor and the scattering rate $\gamma$ are virtually homogeneous on the Fermi surface. 
The momentum-differentiation grows but remains overall small down to $\alpha=0.5$.
Below this mid-point between 3D and 2D, when the Fermi surface gets visibly quasi two-dimensional,
we notice the emergence of  a pronounced momentum-selectivity
in $Z(\vek{k}_F)$ and $\gamma(\vek{k}_F)$. Akin to hole-doped cuprates, excitations at the anti-node (($\pi,0,k_z$)-direction and equivalent) become more short-lived than at the node ($\pi,\pi,k_z$).
Recently, it has been suggested that such pseudogap physics
is intimately linked to the proximity to a van-Hove singularity\cite{wu2019PG}.%
\footnote{See Ref.~\onlinecite{PhysRevX.8.021048} for a detailed analysis of connections between pseudogap physics and the topology of the Fermi surface in 2D.}
In our model, such a feature exists for $\alpha=0$ and is located at half-filling. Regarding the trend with doping, 
the variance throughout the Brillouin-zone is indeed largest for the smallest doping. However, as far as the momentum-selectivity of 
the scattering rate $\gamma$ and the 
quasi-particle lifetime $\tau=1/(Z\gamma)$
on the Fermi surface is concerned, it is largest for the doping that places the chemical potential furthest away from the van-Hove singularity.

Non-local renormalizations also deform the Fermi surface.
Here (left column in \fref{3DFS}), we show the pertinent $\Re\Sigma(\vek{k}_F,\omega=0)$ with respect to $\overline{\Re\Sigma}^{FS} =1/N_{\svek{k}_F}\sum_{\svek{k}_F}\Re\Sigma(\vek{k}_F,\omega=0)$. 
The latter Fermi-surface average can be seen as defining an effective local theory, obtained by a (partial) momentum-projection 
of the D$\Gamma$A self-energy.
Following the arguments detailed in Appendix \aref{sapp1}, the Fermi surface expands
(retracts) where $\Re\Sigma(\vek{k}_F,\omega=0)-\overline{\Re\Sigma}^{FS}>0$ ($<0$). 
In 3D ($\alpha=1$), the Fermi surface is a body with tube-like openings in all axial directions.
Non-local many-body effects shrink the tubes while expanding the inner core where the tubes intersect.
When the hopping in $z$-direction decreases ($\alpha\leq 0.5$), the static non-local shifts of $\Re\Sigma$ expand the Fermi surface near the {\sl xy}-basal 
plane, but narrow the tubal forms oriented along the z-axis, thus magnifying the effective dimensional reduction for correlated electrons.
 
\paragraph{Renormalizations at finite energies.}
The fact that the self-energy coefficients have a larger variation over the entire Brillouin zone than over the Fermi surface ($\Delta^{loc}_{k}a(\vek{k})>\Delta^{FS}_{k}a(\vek{k})$) suggests that non-local effects are larger for excitations at finite energies, $|\epsilon_\svek{k}|>0$.
For $\Re\Sigma(\vek{k},\omega=0)$ this can be rationalized as follows (using for simplicity a static self-energy): The momentum-derivative of $\Re\Sigma(\vek{k})$ at $\svek{k}_F$ reduces the effective mass (see, e.g., Eq.~(1) in Ref.~\onlinecite{jmt_dga3d}). 
Away from the Fermi surface to first order
$\Re\Sigma(\svek{k})\approx (\vek{k}-\vek{k}_F)\cdot \nabla_{\svek{k}}\Re\Sigma(\svek{k})_{\svek{k}=\svek{k_F}}$.
Assuming the derivative / the effective mass to be roughly constant, this effect 
 increases the band-width\cite{jmt_pnict,PhysRevB.87.115110,jmt_svo_extended} via a shift  
 $\Re\Sigma(\vek{k})-\mu{> \atop <}0$ (for $\epsilon_\svek{k}{> \atop <}0$)
 that grows (linearly) with the distance to the Fermi surface. 
 In this sense of widening dispersions, non-local correlations counteract local (dynamical) correlations.
            
\section{Perspective and  Conclusions}\label{perspective}

When tackling a many-body problem---a model Hamiltonian or a strongly correlated material---knowledge about strength, nature and structure of correlation effects is a prerequisite for choosing the most efficient, yet adequate methodology:
Does the system allow for a perturbative\cite{ferdi_gw,RevModPhys.74.601} or weak-coupling\cite{RevModPhys.84.299,Hille2020} treatment? 
Is the low-energy physics known to be dominated by a single microscopic process or fluctuation channel\cite{PhysRevLett.114.236402}? 
Is potential pseudogap physics mostly
driven by short-range fluctuations\cite{ANDERSON1196,Sordi2012}?
Does the self-energy verify certain properties: Is it (to a sufficient degree) local (\`a la DMFT\cite{PhysRevLett.62.324,bible}), static (\`a la Hartree-Fock, DFT+U\cite{PhysRevB.44.943}), or does it approximately obey space-time separation\cite{jmt_dga3d}, \eref{eqn:separation}:
$ 
  \Sigma(\vek{k},\omega)=\Sigma_{\text{static}}(\vek{k})+\Sigma_{\text{local}}(\omega) 
$? 
The validity of \eref{eqn:separation} in 3D for the Hubbard model\cite{jmt_dga3d} and
isotropic cubic materials\cite{Anna_ADGA}, 
advocates that the realistic methodology
DFT+DMFT (that combines DMFT with density functional theory)
can be vastly improved by supplying it with an adequate static but non-local
potential (a $\Sigma_{\text{static}}(\vek{k})$ beyond DFT). The latter highlights the
merits of approaches such as {\it GW}+DMFT\cite{PhysRevLett.90.086402}$^,$\cite{0953-8984-28-38-383001,Tomczak2017review},
QS{\it GW}+DMFT\cite{jmt_sces14,Choi2016}, {SEx+DMFT}\cite{paris_sex}, 
or space-time separated {\it GW}\cite{jmt_dga3d},
in all of which the non-local contribution to the self-energy (in the Wannier sense) is empirically
found to be essentially static
at least at low energies.%
\footnote{\eref{eqn:separation} may also lead to
simplifications in the dual fermion approach\cite{PhysRevB.92.144409}.}

As a consequence, it is valuable to know the limits of validity of \eref{eqn:separation}.
Our results suggest that the error incurred by assuming space-time separability
of the self-energy becomes prohibitive at an anisotropy corresponding to about half-way ($\alpha\approx 0.5$) between the isotropic 3D and the planar 2D case.

Our findings hence support the view (the expectation) that in cuprates (d$^9$ nickelates\cite{LaNiO2,Li2019,Wilson2019}) no local theory can account for the rich many-body physics at play. 
Clearly, in 2D the self-energy does not obey space-time separation. 
At least at half-filling other simplifications may, however, apply for the Hubbard model\cite{PhysRevB.93.195134}.

The iron pnictides and chalcogenides reside in an $\alpha$-regime that 
suggests the presence of sizable non-local renormalizations, see \fref{fig:survey}.
Concurringly, diagrammatic fluctuation techniques recently
demonstrated\cite{PhysRevLett.123.256401,bhattacharyya2020nonlocal}
that {\it static} non-local renormalizations are indeed non-negligible---in line with previous experimental analyses \cite{PhysRevLett.110.167002}.
Non-local corrections to {\it dynamical} renormalizations, however, were previously shown\cite{PhysRevB.95.195115} to be small, at least away from the systems' N{\'e}el
temperatures.
These recent results
substantiate earlier claims based on perturbation theory, and strengthen the space-time separability of the self-energy in the iron pnictides and chalcogenides proposed in Refs.~\onlinecite{jmt_pnict,jmt_sces14} 
\footnote{Note that our study does not make any statements about non-local self-energies drawing from 
non-local (beyond-Hubbard model) interactions. The latter can be notable\cite{PhysRevB.95.245130} irrespective of the crystal-structure. In fact, non-local exchange self-energies are sizable in
pnictides\cite{jmt_pnict,jmt_sces14,paris_sex}, as well as in isotropic correlated oxides\cite{jmt_svo_extended,PhysRevB.87.115110,0295-5075-108-5-57003}. In a perturbative treatment, these non-local self-energy contributions are static, thus verifying \eref{eqn:separation}.
Also, local {\it multi-orbital} interactions
that can lead to non-local correlations\cite{PhysRevB.91.235107} are (by construction) not included here.}%
.

According to the anisotropy classification of materials in \fref{fig:survey},
we ascertain that perovskite oxides are firmly in the realm of \eref{eqn:separation}: {\it dynamical} self-energies will be essentially local.
However, we found {\it static} non-local renormalizations (beyond DMFT) to be non-negligible 
also for cubic systems---motivating the above approaches to include a $\Sigma_{\text{static}}(\vek{k})$ (beyond DFT).
Yet, most importantly, our findings strongly suggest that geometric constraints in (simulations of) oxide-based heterostructures\cite{doi:10.1146/annurev-matsci-070115-032057,Janson2018,Lechermann2018} or ultra-thin films\cite{Potthoff1999,Liang_SRO,Golalikhani2018,PhysRevB.97.075107,Matthias_SVOultra,loon2020coulomb}
may results in strong {\it dynamical} non-local effects---posing in particular limits on the applicability of DMFT.
We believe that our anisotropy measure of \eref{alpha} can provide guidance 
to identify such systems prior to actual many-body calculations.
Moreover, our findings call for reexamining layered systems such as sodium cobaltates and 124-ruthenates with realistic many-body approaches beyond DMFT\cite{Anna_ADGA,Tomczak2017review,RevModPhys.90.025003}.

\section{Acknowledgements}
The authors gratefully acknowledge discussions with D.\ Vollhardt.
The present work was supported by the Austrian Science Fund (FWF) through the Erwin-Schr\"odinger Fellowship J 4266 - ``{\sl Superconductivity in the 
vicinity of Mott insulators}'' (SuMo, T.S.),
project ``{\sl Simulating Transport Properties of Correlated Materials}'' (LinReTraCe P~30213-N36, J.M.T.), and
 project ``{\sl Merging dynamical mean-field theory and functional renormalization group}'' (I 2794-N35, A.T.),
as well as the European Research Council, with Grant No. 319286 (QMAC, T.S.) and Grant No. 725521 (TheOne, B.K.).
Calculations were partially performed on the Vienna Scientific Cluster (VSC).

\appendix
\section{Details on the dynamical vertex approximation calculations}
\label{app:dga}
Diagrammatically the approximation made in DMFT is assuming the one-particle irreducible self-energy to be purely local also in finite-dimensional systems: $\Sigma(\mathbf{k},i\omega_n)\!\approx\!\Sigma(i\omega_n)$. The dynamical vertex approximation (D$\Gamma$A)\cite{toschi_dynamical_2007,RevModPhys.90.025003} raises this assumption to the two-particle analog of the self-energy, the fully irreducible two-fermion scattering vertex $\Lambda_{\Omega\omega\omega'}^{\mathbf{q}\mathbf{k}\mathbf{k}'}\!\approx\!\Lambda_{\Omega\omega\omega'}$. This leads to a systematic inclusion of non-local correlations on every length scale. 
Without an {\it a priori} knowledge of which physical scattering channel dominates the physics\cite{PhysRevLett.114.236402,Gunnarsson2016}, for D$\Gamma$A calculations the (computationally very demanding) parquet equations have to be solved self-consistently. 
In this work,  we restricted ourselves to non-local correlations stemming from magnetism. Then the D$\Gamma$A equations can be considerably simplified, since the full parquet D$\Gamma$A can be restricted to its single-shot ladder (Bethe-Salpeter) version with Moriyaesque $\lambda$-corrections in both the charge and the spin channel channel [Eq.~(6) in  Ref.~\onlinecite{rohringer_impact_2016}]. 
We used the code available at Ref.~\onlinecite{ladderdgacode}. 
We obtained the necessary two-particle 
Green's function after a self-consistent DMFT calculation from an exact diagonalization impurity solver (with four bath sites), whose results we carefully checked against the ones from continuous time quantum Monte Carlo \cite{w2dynamics}. 
For the Bethe-Salpeter ladders and Dyson-Schwinger equation we used a momentum grid with a maximum linear mesh sizes of $N_{q}\!=\!60$ and $N_{k}=20$ and the total number of fermionic as well as bosonic Matsubara frequencies being $N_{i\omega}\!=\!N_{i\Omega}\!=\!120$. 
We calculated the self-energy for $N_{k,\Sigma}=2601$ points in the Brillouin zone.
\section{Effects of $\Re\Sigma(\vek{k}_F,\omega=0)$}\label{sapp1}
The Fermi surface (FS) using the FS-averaged D$\Gamma$A self-energy,
$\overline{\Re\Sigma}^{FS}=1/N_{\svek{k}_F}\sum_{\svek{k}_F}\Re\Sigma(\vek{k}_F,\omega=0)$, is given by
\begin{equation}
\epsilon_{\svek{k}_F^0}+\overline{\Re\Sigma}^{FS}=0.\label{app1}
\end{equation}
Here and in the following we shall always absorb the chemical potential $\mu$ in $\Re\Sigma$.
Since FS-averaged D$\Gamma$A defines a local theory, the Fermi surface is by construction the same as within DMFT.
The Fermi surface of the full D$\Gamma$A solution is obtained as:
\begin{eqnarray}
\epsilon_{\svek{k}_F}+\Re\Sigma(\vek{k}_F)&=&0\\
\epsilon_{\svek{k}_F^0}+\Re\Sigma(\vek{k}_F^0)+\delta_k\,\partial_k\left(\epsilon_k+\Re\Sigma(k)\right)_{k=\svek{k}_F^0}&=&0
\end{eqnarray}
Where we linearized around the above $\vek{k}_F^0$: $\delta_k=\vek{k}_F-\vek{k}_F^0$ (and, for keeping the notation light, omitted complications introduced by the directional dependence of the momentum-derivatives).
Now we introduce $\Delta \Re\Sigma(\vek{k}_F^0)=\Re\Sigma(\vek{k}_F^0)-\overline{\Re\Sigma}^{FS}$ and use \eref{app1}, then:
\begin{eqnarray}
\Delta\Re\Sigma(\vek{k}_F^0)+\delta_k\times \partial_k\left(\epsilon_\svek{k}+\Re\Sigma(\vek{k})\right)_{k=\svek{k}_F^0}&=&0
\end{eqnarray}
{yielding the Fermi-surface deformation}
\begin{eqnarray}
    \delta_k&=&\Delta\Re\Sigma(\vek{k}_F^0) \times \frac{m^*}{m}\frac{Z}{v_{\svek{k}_F^0}}
\end{eqnarray}
where we used the definition of the effective mass (e.g., Eq.~(1) in Ref.~\onlinecite{jmt_dga3d}).
We find $\forall \alpha\forall \vek{k}\in \hbox{FS}$:  $\hat{\vek{e}}_\svek{k}\cdot\nabla_{\svek{k}}\epsilon(\vek{k})>0$ as expected for hole-doping.
Hence the sign of $\Delta\Re\Sigma(\vek{k}_F^0)$ determines whether for that $\vek{k}$ the FS expands ($+$) or shrinks ($-$).

\begin{figure}[h!]
            \centering 
            \vspace{0.5cm}
            \includegraphics[width=1.\textwidth]{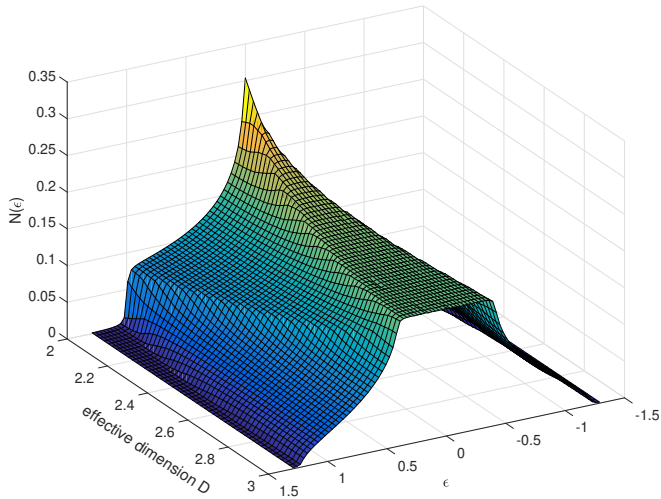}
            \caption{Density of states $N(\epsilon)$ of the tetragonal dispersion of \protect\eref{dispersion} for the dimension D varying continuously between 3D ($\alpha=1$) and 2D ($\alpha=0$).  
                        }
            \label{DOS}
        \end{figure}


%

\end{document}

%% file: Header_main.tex
\usepackage[pdftex]{graphicx}   

\usepackage{caption}
\usepackage{subcaption}

\usepackage{epstopdf}
\usepackage{floatrow}
\newfloatcommand{capbtabbox}{table}[][\FBwidth]

\usepackage[utf8]{inputenc} 
\usepackage{csquotes} 
\usepackage{paralist}

\usepackage[utf8]{inputenc}
\usepackage{hyperref}                                                   
\hypersetup{
  colorlinks   = true, 
  urlcolor     = cyan, 
  linkcolor    = blue, 
  citecolor    = blue  
}

\newcommand{\kvec}[0]{\mathbf{k}}

\newcommand{\DfsRe}[0]{$\Delta_{\kvec}^{FS}\text{Re}(\Sigma)$}

\newcommand{\minmaxRe}[0]{$\delta^{FS}\text{Re}(\Sigma)$}